\documentclass[journal=jacsat,manuscript=article]{achemso}
\usepackage{xcolor}
\usepackage[table]{xcolor}
\usepackage[normalem]{ulem}
\usepackage[version=3]{mhchem} 
\usepackage{float}

\author{Elena Pinilla-Cienfuegos}
\affiliation{Nanophotonics Technology Center, Universitat Politècnica de València, Valencia E46022, Spain}
\email{epinilla@ntc.upv.es}

\author{Lucas Mascaró-Burguera}
\affiliation{Nanophotonics Technology Center, Universitat Politècnica de València, Valencia E46022, Spain}

\author{Ramón Torres-Cavanillas}
\affiliation{Instituto de Ciencia Molecular, Universitat de València, Valencia, 46980 Spain}

\author{J. Ignacio Echavarría}
\affiliation{Department of Optics, Faculty of Physics, University Complutense of Madrid. Plaza de Ciencias 1, 28040 Madrid, Spain}

\author{Alejandro Regueiro}
\affiliation{Instituto de Ciencia Molecular, Universitat de València, Valencia, 46980 Spain}

\author{Eugenio Coronado}
\affiliation{Instituto de Ciencia Molecular, Universitat de València, Valencia, 46980 Spain}

\author{Javier Hernandez-Rueda}
\affiliation{Department of Optics, Faculty of Physics, University Complutense of Madrid. Plaza de Ciencias 1, 28040 Madrid, Spain}
\email{fj.hernandez.rueda@ucm.es}

\title[An \textsf{achemso} demo]
 {Unveiling Spin Transition at Single Particle Level in Levitating Spin Crossover Nanoparticles}


\begin{document}

\begin{tocentry}
\includegraphics[width=8.3 cm]{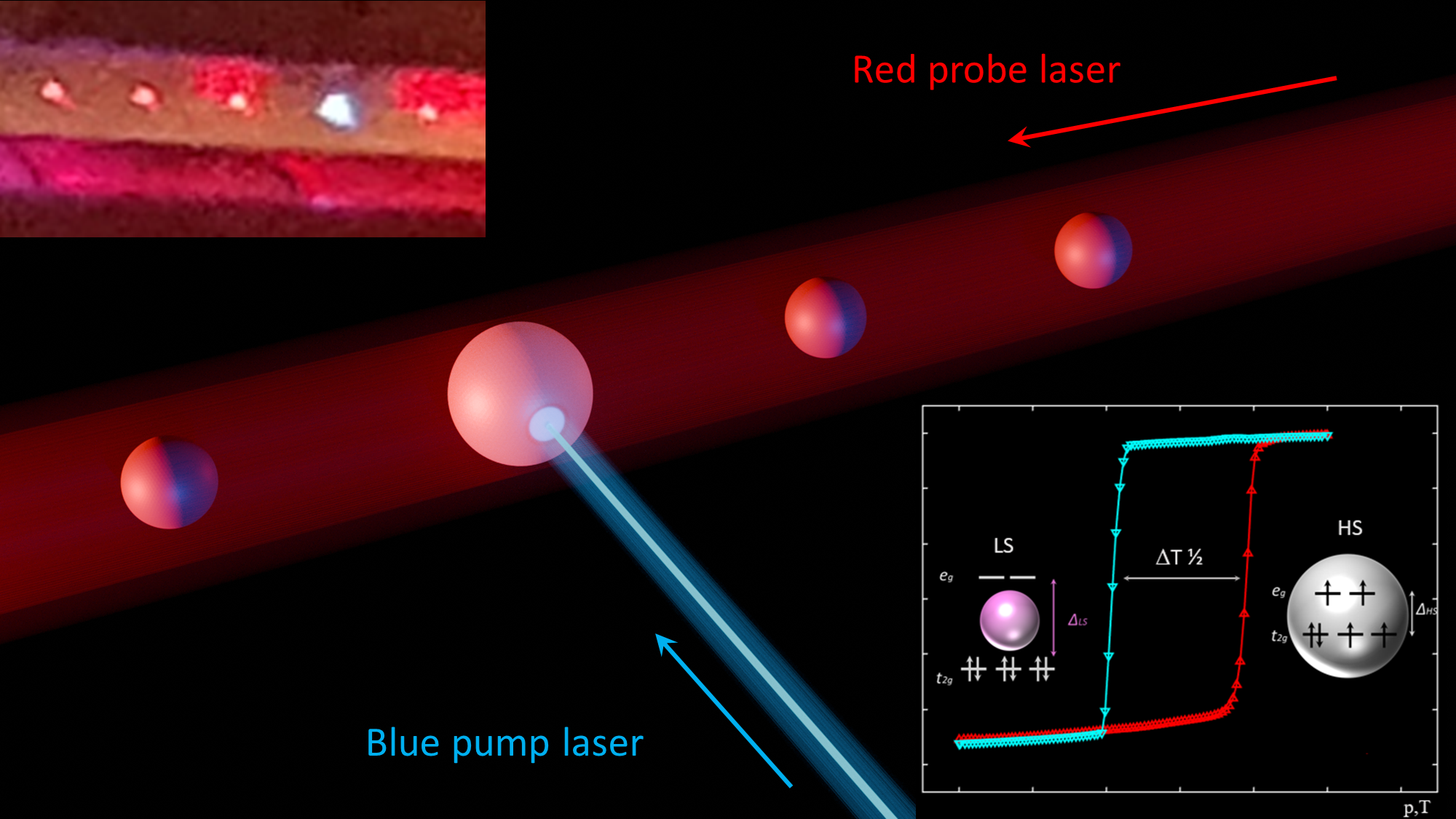}
\end{tocentry}

\begin{abstract}
The ability to control and understand the phase transitions of individual nanoscale building blocks is key to advancing the next generation of low-power reconfigurable nanophotonic devices. To address this critical challenge, molecular nanoparticles (NPs) exhibiting a spin crossover (SCO) phenomenon are trapped by coupling a quadrupole Paul trap with a multi-spectral polarization-resolved scattering microscope. This contact-free platform simultaneously confines, optically excites, and monitors the spin transition in Fe(II)–triazole NPs in a pressure-tunable environment, eliminating substrate artifacts. Thus, we show light-driven manipulation of the spin transition in levitating NPs free from substrate-induced effects. Using the robust spin bistability near room-temperature of our SCO system, we quantify reversible opto-volumetric changes of up to 6\,$\%$, revealing precise switching thresholds at the single-particle level. Independent pressure modulation produces a comparable size increase, confirming mechanical control over the same bistable transition. These results constitute full real-time control and readout of spin states in levitating SCO NPs, charting a route toward their integration into ultralow-power optical switches, data-storage elements, and nanoscale sensors.  

\end{abstract}

\section*{Introduction}
Reconfigurable nanophotonic devices rely on phase-change materials (PCMs) to dynamically control light \cite{Wuttig2017}. To be suitable for practical implementation in integrated photonic circuits and optoelectronic systems, PCMs must meet key requirements such as high optical contrast, fast and reversible switching, low energy consumption, and scalability.\cite{Parra2021} While conventional PCMs such as chalcogenides and vanadium dioxide satisfy some of these requirements, they are often limited by high switching energy, optical losses, and, in certain cases, poor cycling stability, which hinders their performance and long-term reliability in photonic applications.\cite{ Ko2022, Briggs2010, hernandez2011coherent}.
Molecular PCMs such as spin crossover (SCO) materials stand out as particularly promising candidates, since their properties can be chemically tailored for advanced photonic functionalities. 
\cite{Gutlich2004, Ridier2025, Zhang2024SCOswith} These compounds, generally based on octahedral Fe(II) coordination complexes, reversibly switch between two electronic configurations, the so-called high spin state (HS) and low spin state (LS), in response to external stimuli, such as temperature, light irradiation, or pressure variations\cite{D3SC01495A,https://doi.org/10.1002/adfm.202000447,doi:10.1021/acs.inorgchem.4c02170,torres2019downsizing,GavaraEdo2023,sanchis2021plasmon,Remili2025}. This transition is accompanied by substantial changes in structural (volume), magnetic, optical, electrical, mechanical and thermal properties that become evident to the naked eye through changes in color.\cite{Gutlich2004,Epinilla2010,Yang2024,regueiro2024controlling} Among these changes, the optical response is particularly relevant for photonic integration: the spin transition induces a measurable change in the refractive index, predicting a $\Delta n \approx$ 0.01–0.1 across the whole vis–NIR range when going from the HS to the LS state. Equally significant is the accompanying reversible volume increase, often of the order of several percent in volume ($\Delta V \approx$ 6 \,$\%$ in typical Fe(II) complexes) arising from the structural rearrangement between the high and low-spin configurations\cite{Molnr2017}. For these reasons, and since SCO is a molecular phenomenon that can function even at the single-molecule scale, the scientific community has focused in recent years on the development of SCO systems at the nanoscale for their integration into nanodevices.\cite{zhang2024spin,Molnr2017,dugay2019sensing} Notably, in many SCO compounds, intermolecular interactions between the molecular metal complexes give rise to a cooperative spin transition with thermal hysteresis, creating a bistable temperature window in which either spin state may be stabilized, a key feature for applications in memory storage and switching technologies\cite{https://doi.org/10.1002/adfm.202102469,Yu2025,Bousseksou2011}.

However, manipulation and readout of individual SCO NPs have proven to be a major challenge. Previous studies employing ultrafast electron microscopy to probe mechanically induced or plasmon-assisted spin transitions in supported NPs have provided significant insight, but are inherently limited by substrate-induced effects that can interfere with accurate measurement of their intrinsic properties \cite{hu2023laser, hu2021photo, mba2023lattice}. This challenge becomes even more pronounced when attempting to integrate individual SCO NPs into functional nanodevices, as illustrated by a notable work that involved the integration of ca. 10\,nm NPs of the polymeric chain compound [Fe(HTrz)2(Trz) (BF$_4$)] (where Trz refers to the triazole ligand) between two gold electrodes.\cite{prins2011room} Interestingly, thermal bistability in transport properties was detected near room temperature, which was associated with the spin transition. However, this device lacked sufficient stability and reproducibility for practical applications. As a consequence, most reported works have relied on researchers using compressed NP powders or embedding particles within polymer matrices or depositing them on conducting two-dimensional systems.\cite{Porel2004, Salmon2018,nunez2023hybrid,Dugay2015,torres2021spin} Some efforts have involved the use of soft lithography to position NP assemblies (rather than single NPs) of the aforementioned compound between gold electrodes.\cite{torres2024bistable} An even more sophisticated procedure takes advantage of the anisotropy of large SCO microrods to trap them between electrodes using dielectrophoresis.\cite{Lefter2015,Rotaru2013} Unfortunately, these approaches do not allow for precise control of individual SCO nanosystems and typically result in assemblies of nano/microparticles within the device that, although stable, still lack reproducibility and uniformity. In this context, we introduce a novel approach based on the use of a quadrupole Paul trap to isolate SCO NPs in a controlled atmosphere. This platform enables substrate-free, real-time optical control, and monitoring of SCO transitions at the single-particle level, with environmental control over pressure, temperature, and laser intensity, while fully eliminating substrate-induced effects. A Paul trap uses an oscillating electric field to confine ions or single NPs in a three-dimensional space.\cite{https://doi.org/10.1002/anie.199007391,https://doi.org/10.1002/rcm.607,conangla2019optimal,10.1063/1.4893575} In recent years, quadrupole traps have been shown to be ideal for isolating single NPs, enabling precise spectroscopic studies in a controlled environment.\cite{Hernandez-Rueda:19,PhysRevResearch.3.013018,maestrohernandez2025} More recently, Paul traps have been used to manipulate single NPs to investigate macroscopic quantum states, leading to new insights with great potential for sensing applications with unprecedented sensitivity.\cite{doi:10.1126/science.abg3027,tebbenjohanns2021quantum,piotrowski2023simultaneous, carlon2025motional,rossi2024quantum,sund2024simulation}

In this work, we have used this procedure to isolate SCO NPs of the amino derivative of the Fe(II)–triazole compound [Fe(NH$_2$trz)3(NO$_3$)$_2$]. We selected this system because of its robustness, sharp hysteretic response, and spin transition centered close to room temperature, which represents an ideal candidate for the design of low-power bistable devices.
The trap setup combined with a multi-spectral polarization-resolved scattering microscope allowed us to simultaneously confine, excite, and probe a spin transition at the single NP level. The transition between spin states was triggered using a tunable external pump laser, where the transition is thermally induced by adjusting the laser power. Because the trap operates in a vacuum chamber to ensure stable confinement of NPs, each levitates freely in the Paul trap so that its full three-dimensional size change can be measured without substrate constraints. Moreover, the precise pressure control of the vacuum chamber allows us to actively modulate that size and keep track of pressure-induced volumetric changes in real time. 
By analyzing the scattering signal and size variations, we were able to identify the spin state of each particle and pinpoint the specific excitation conditions of the laser power at which the spin transition occurs, along with the corresponding size changes induced by vacuum tuning. Lastly, we discuss the potential of this new phase-change material for developing low-power nanophotonic switches and detectors.

\section*{RESULTS AND DISCUSSION}

\subsection*{Ex-situ characterization of SCO NPs}
\begin{figure}[H]
\centering
\includegraphics[width=1\linewidth]{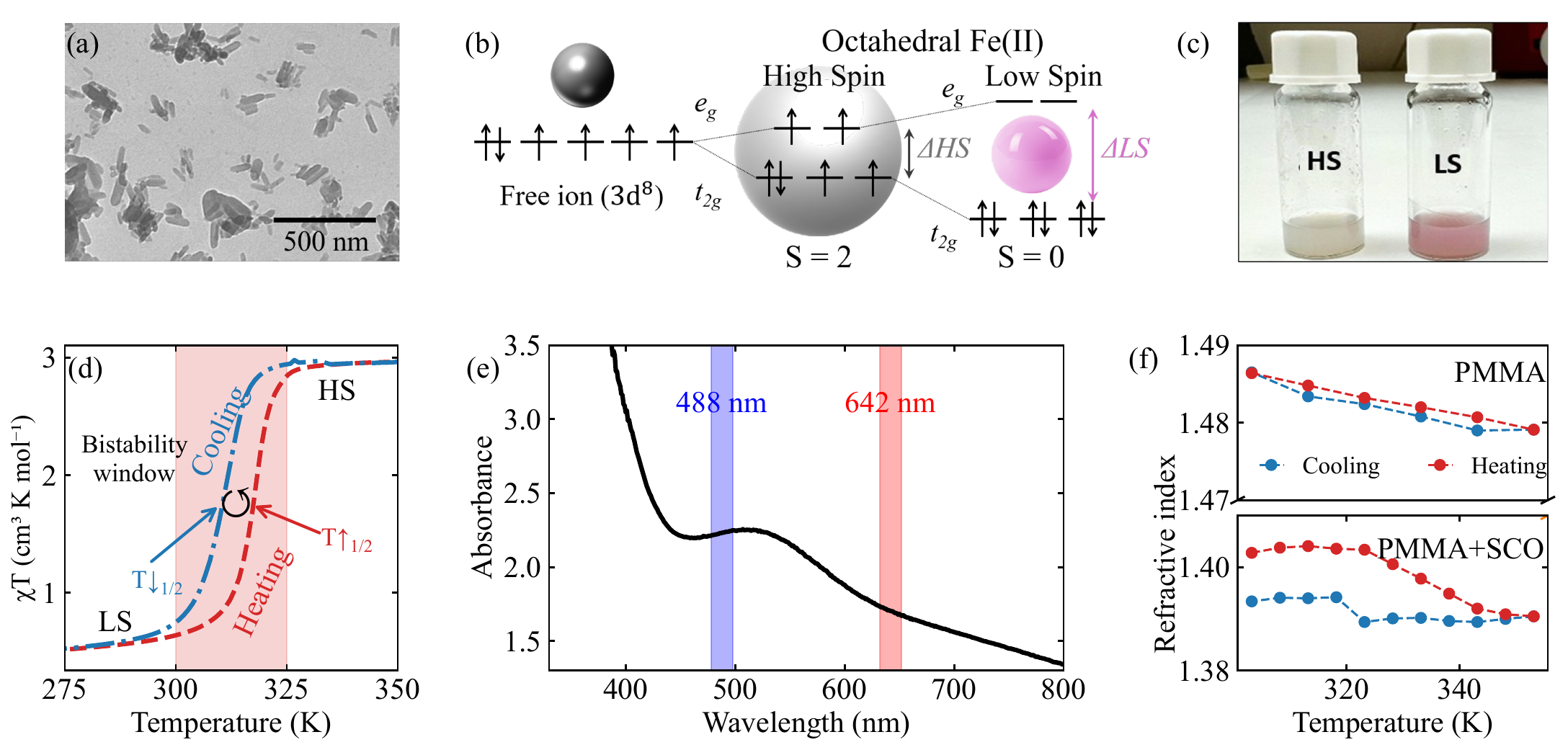}
\caption{(a) Representative TEM image of Fe(NH$_2$trz)3(NO$_3$)$_2$ particles. (b) Schematic of the electron redistribution between the LS and HS configurations in an octahedral Fe(II) coordination compound upon different external stimuli (pressure, temperature and light irradiation). (c) Picture of a bistable SCO/PMMA solution at room temperature, with low (pink, LS) and high (white, HS) spin states. (d) Heating-and-cooling cycles of $\chi T$ as a function of temperature. (e) Absorption spectrum of NPs redispersed in ethanol. (f) Ellipsometry measurements with heating-and-cooling cycles of the real part of the refractive index for a bare PMMA layer (top graph) and SCO/PMMA film (bottom graph).}
\label{fig1}
\end{figure}

Fe(NH$_2$trz)3(NO$_3$)$_2$ exhibits a spin transition near room temperature, which allows low-energy activation. Since it contains multiple anions (NO$_3$$^-$) in their structure, this compound is well-suited for the formation of partially charged NPs under applied voltage, which is a key requirement to effectively trapping NPs (see Fig.~S1). NP synthesis was carried out using a reverse micelle protocol, which involved mixing two separate microemulsions, as described elsewhere.\cite{regueiro2024unlocking} We adapt the experimental conditions to synthesize NPs with sizes that range between 100\,nm and 300\,nm (see Experimental Methods), which was confirmed by dynamic light scattering (DLS, in Fig.~S2) and by transmission electron microscopy (TEM, in Fig.~\ref{fig1}(a)).

We confirmed that the compound was successfully formed using infrared spectroscopy (IR) and X-ray diffraction (XRD) (Fig.~S3). The energy diagrams presented in Fig.~\ref{fig1}(b) illustrate the HS and LS configurations for a Fe(II) complex, which for the HS state features four unpaired electrons with S = 2 that yield a strongly paramagnetic configuration. Its LS state counterpart has no unpaired electrons and is diamagnetic. This fundamental difference underlies the SCO behavior, which can be triggered by external stimuli such as variations in temperature, pressure, or light intensity, as mentioned in the introduction. The distinct number of unpaired electrons in the high- and low-spin states results in a marked change in color and magnetic moment. Fig.~\ref{fig1}(c) shows a picture of a bistable SCO NP suspension in a poly(methyl methacrylate) (PMMA) matrix at RT, whose color changes from typical pink at LS to milky-white at HS after a heating cycle due to a spin transition. Therefore, to accurately determine the transition temperature of the synthesized NPs, we investigated their magnetic response by measuring the thermal dependence of  $\chi T$, in cm$^3$K mol$^{-1}$. Fig.~\ref{fig1}(d) shows a well-defined thermal spin transition from the diamagnetic LS (S = 0) state to the paramagnetic HS (S = 2) state around 320 K (T$\uparrow$1/2) upon heating. During cooling, the LS state is recovered below 310 K (T$\downarrow$1/2), demonstrating a reversible spin transition with a bistable window of approximately $\Delta T$  =  10 K. This hysteretic behavior is consistent across successive thermal cycles.

To optically characterize Fe(NH$_2$trz)3(NO$_3$)$_2$ NPs, we performed UV-Vis spectroscopy and ellipsometry measurements, as shown in Figs.~\ref{fig1}(e) and (f), respectively. The particles were embedded in PMMA to enable their optical characterization. The room temperature UV-Vis absorption spectrum of NPs redispersed in ethanol exhibits the characteristic LS state signature, marked by a broad absorption band centered at 520\,nm with a bandwidth of 100\,nm. This feature gives the LS state its typical pink color. In contrast, this absorption band disappears in the HS state, rendering the dispersion milky-white, as shown in Fig. 1(c). Based on this spectrum, we selected a 488\,nm blue laser to trigger the spin transition and a 642\,nm red laser for scattering-based size measurements of individual particles. For spectroscopic ellipsometry measurements, the resulting nanocomposite (SCO/PMMA) was spin-coated onto a silicon substrate, producing a thin and optically smooth layer suitable for ellipsometric analysis. This approach allowed us to extract the complex refractive index of the Fe(NH$_2$trz)3(NO$_3$)$_2$ compound and monitor its variation across the spin transition as a function of temperature. The graphs in Fig.~\ref{fig1}(f) illustrate the real part of the refractive index during the heating and cooling cycles for the SCO/PMMA film (bottom panel) and for a bare PMMA layer (top panel) at a wavelength of 642\,nm. Similarly to $\chi T$, the refractive index presents a thermal hysteresis, while the PMMA film presents a linear behavior. The measurements revealed a refractive index of $n_{LS}$ = 1.4028 for the LS state and $n_{HS}$ = 1.3905 for the HS state, corresponding to an optical contrast of $\Delta n$ = 0.0123 across the spin transition (see full ellipsometry characterization in the Supporting Information Fig.~S5).

\subsection*{Trapping levitating SCO NPs.}

\begin{figure}[H]
\centering
\includegraphics[width=\linewidth]{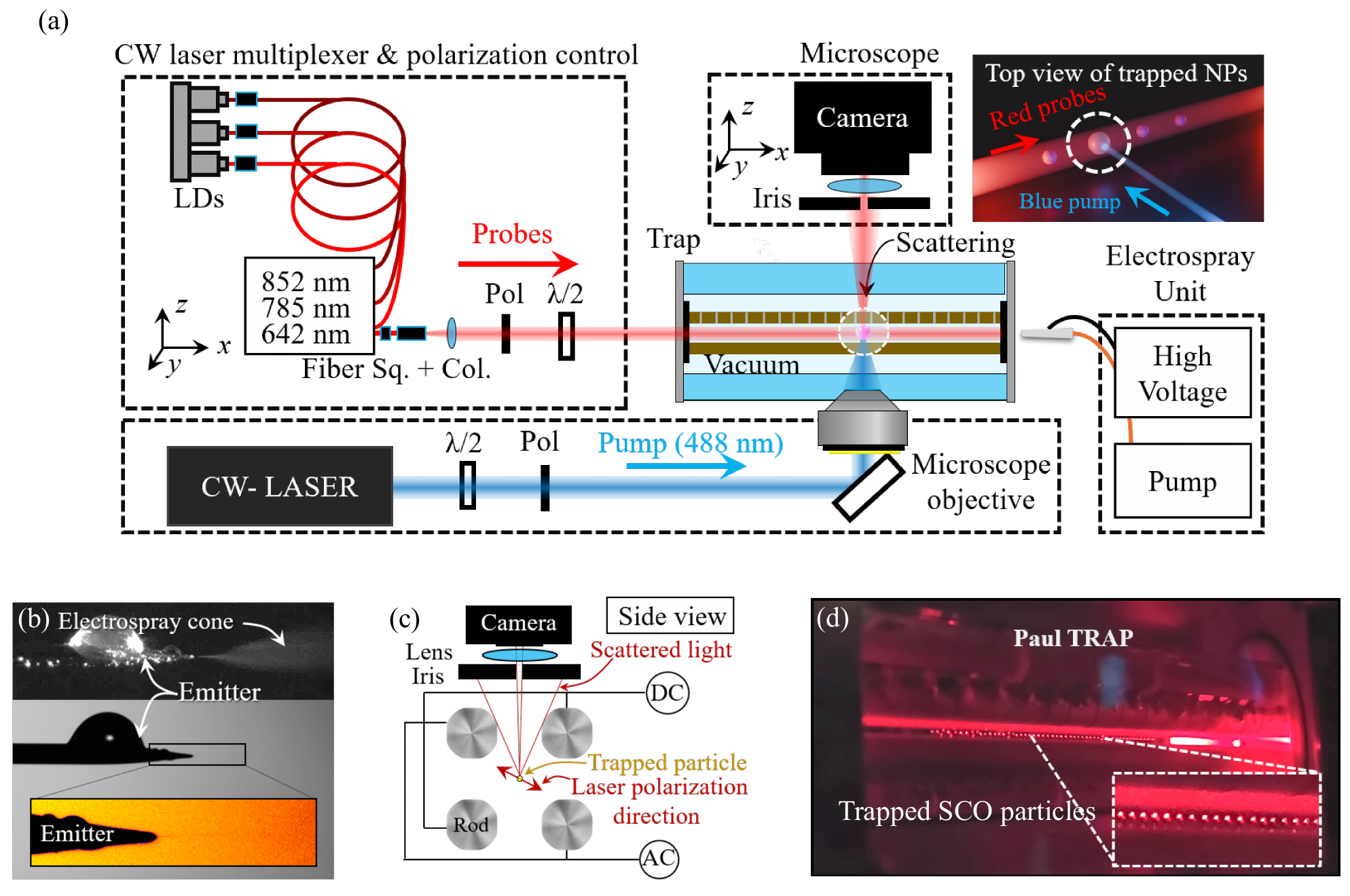}
\caption{Experimental setup for polarization-resolved scattering measurements on levitated SCO nanoparticles.  
(a) Schematic of the optical system: three continuous-wave (CW) probe lasers at 642, 785, and 852~nm (LDs) are combined via a fiber multiplexer, collimated, and polarization-controlled using a half-wave plate ($\lambda/2$) and linear polarizer (Pol) before entering the trap along the $x$-axis. A separate CW laser at 488~nm is used as a pump beam when required. The trap is placed inside a vacuum chamber with optical access through a microscope objective, and the scattered light is collected along the $y$-axis and imaged onto a camera. Inset: top-view 3D sketch image of trapped nanoparticles under simultaneous red probe and blue pump illumination.  
(b) Electrospray source used to inject SCO NPs into the trap. Dark-field and shadowgraphy microscopy images showing the emitter and Taylor cone formation.  
(c) Side-view schematic of the linear Paul trap indicating the positions of the DC endcap electrodes and AC rods, as well as the polarization direction of the probe laser and the scattered light collection geometry.  
(d) Photograph of the Paul trap during operation, showing the trapped SCO NPs illuminated by the probe beams.
 }
\label{fig2}
\end{figure}

To investigate the spin transition at the single NP level, we make use of a Paul trap embedded in a vacuum chamber and monitor its response to external stimuli using multi-spectral polarization-resolved scattering microscopy (Fig.~\ref{fig2} (a)).\cite{Hernandez-Rueda:19,maestrohernandez2025} 
Continuous-wave probe beams at 642, 785, and 852~nm are combined in a fiber multiplexer, collimated, polarization-controlled, and aligned collinearly to the trap axis. A separate 488~nm CW laser serves as a pump for optical excitation. Scattered light from the trapped particle is collected perpendicularly to the trap axis by a long-working-distance microscope objective and imaged onto a camera for polarization-resolved scattering measurements (see Methods section for details). The inset on the upper right displays a 3D sketch image of trapped NPs, highlighting the co-linear red probe beams and the perpendicular blue pump beam. This configuration enables simultaneous optical excitation and multi-wavelength, polarization-resolved optical detection of single particles while varying environmental parameters such as chamber pressure. We introduce Fe(NH$_2$trz)3(NO$_3$)$_2$ SCO NPs (hereafter SCO NPs) in the trap by using an electro-spray system that combines a syringe with an ethanol solution of the NPs coupled to an emitter connected to a high voltage ($\approx$ 3 kV). The trap entrance is grounded, thus the voltage difference generates a Taylor cone, where particles are separated and propelled towards the trap input port as shown by dark-field and shadowgraphy microscopy in Fig.~\ref{fig2}(b).\cite{gaskell1997electrospray, alda2020levitodynamics} The trap consists of four metallic rods diagonally connected to either an AC sinusoidal signal or a constant DC signal as sketched in Fig.~\ref{fig2}(c). This configuration generates a quadrupolar potential able to trap single NPs along the trap axis, which obey the Mathieu equations of motion. The precise choice of frequency and amplitude of the potential offers the flexibility of being able to trap particles with on-demand sizes, i.e. these traps were originally conceived as mass selectors. Thus, we set the trap setup parameters to purposely trap SCO NPs with a size around 300\,nm, which simultaneously offers good trapping stability and a high scattering signal during experiments (see Experimental Methods). This size is compatible with the size distributions measured using TEM (see Fig.~\ref{fig1}(a)) and dynamic light scattering (DLS, Fig.~S2). Our trap setup uses an automated scattering microscope to collect images of isolated SCO NPs that are illuminated with low-intensity linearly polarized laser light at different polarization angles, as shown in Fig.~\ref{fig2}(c). Fig.~\ref{fig2}(d) shows a picture of the Paul trap with an array of SCO NPs levitated in a pressure-controlled environment, which is ideal for studying the spin transition upon laser irradiation and pressure variations.\cite{Hernandez-Rueda:19,maestrohernandez2025} The scattering response of isolated particles aids in gaining insight into how they behave under external stimuli, which leads to a better understanding of the SCO mechanism at the single particle level. This method has great potential to reveal strategies to fine-tune the spin transition for specific applications in the presence of different environments or nanophotonic platforms.

\subsection*{Light-induced reversible spin transition of trapped SCO NPs}

\begin{figure}[H]
\centering
\includegraphics[width=\linewidth]{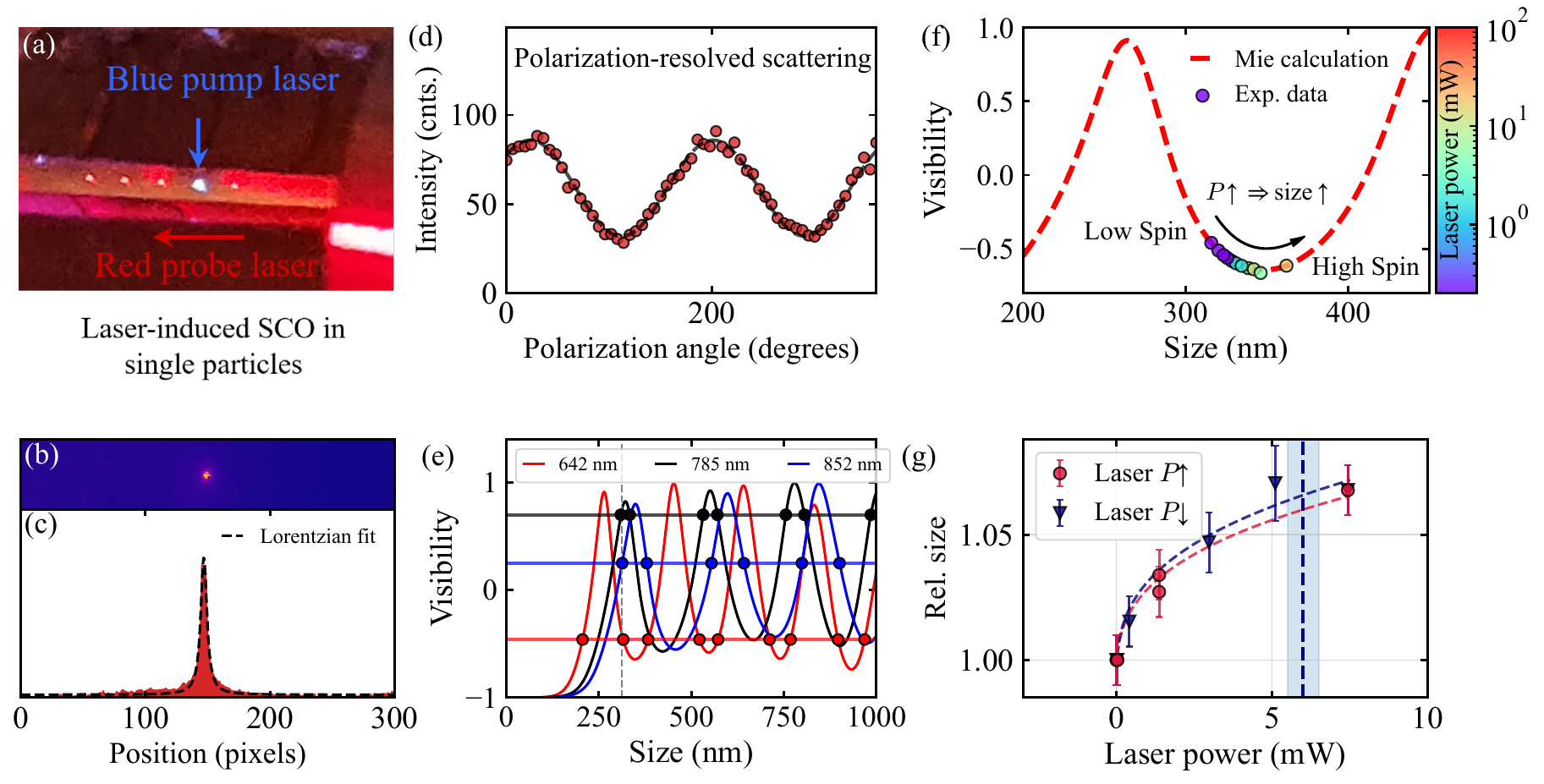}
\caption{(a) Image of an array of isolated SCO NPs inside the trap, which are illuminated with a collimated red probe laser and excited with a focused blue pump laser. (b) Scattering micrograph and its (c) integrated profile of a trapped NP illuminated with a CW-laser at 642\,nm. The dashed-red line illustrates a fit using a Lorentzian function. (d) Scattering intensity as a function of the laser linear polarization angle. The dashed-white line corresponds to a sinusoidal fit. (e) Visibility of SCO NPs as a function of their size at 642\,nm, 785\,nm and 852\,nm laser wavelengths. The solid lines were calculated using Mie theory. The horizontal lines illustrate the experimental visibilities extracted from the data. The markers display the intersections with the Mie visibilities. Dashed vertical line corresponds to a size of 315 nm at zero laser excitation and ambient pressure. (f) Graph of the visibility at 642\,nm versus NP size. These data were measured for three independent trapped SCO NPs irradiated using several laser intensities, which can be found using the color code on the right-hand side. (g) Relative laser-induced size change of trapped SCO NPs as a function of laser excitation intensity at $\lambda =$ 488\,nm. The relative sizes were retrieved following the method illustrated in panel (e). Red circles (P↑) correspond to the up-sweep and blue triangles (P↓) to the down-sweep; dashed lines are guides to the eye. The vertical dashed line marks 6 mW, the power at which independent Raman measurements on non-trapped NPs indicate the HS state (Fig.~S4).}
\label{fig3}
\end{figure}

In the following, we inspect levitating SCO particles using polarization-resolved scattering microscopy to attain key insights into their optical response and size changes upon reversible laser-induced spin transition. During each experimental run, trapped SCO NPs are independently illuminated with two perpendicular continuous wave (CW) laser beams at 488\,nm and 642\,nm as shown in Fig.~\ref{fig3}(a), which are used to excite the particle and probe its scattering response, respectively. The excitation beam at 488\,nm is focused on the NP using a microscope objective (Mitutoyo, 10$\times$, NA=0.28) while the much less intense collimated probe beam at 642\,nm illuminates trapped NPs. 

Using the trap and the in situ microscope shown in Fig.~\ref{fig2}(a), we collect scattering snapshots for polarization angles of the probe laser beam that range from 0 to 2$\pi$. These data provide the scattering response of isolated SCO NPs and, from it, their size under several laser excitation or pressure conditions. Figs.~\ref{fig3}(b) and (c) present a typical scattering microscopy snapshot of a levitating SCO NP and its corresponding integrated signal, where we fit a Lorenztian function to extract a background-free signal.

Fig.~\ref{fig3}(d) presents an example of the scattering response of an isolated SCO NP as a function of the laser polarization angle at 642\,nm. We measure scattering curves at three wavelengths (642\,nm, 785\,nm and 852\,nm) to unambiguously extract the particle size. From the scattering curves, we extract the optical visibilities by fitting sine functions to the data, which are then used to infer the particle size. The relationship between the electric field of the probe laser $E_{in}$ and the scattered field $E_{sc}$ depends on the dielectric function, size and shape of the particle and on the laser wavelength $\lambda$, and is given by:

\begin{equation}
    \binom{E_{\| } ^{sc}}{E_{\perp } ^{sc}}=\frac{\lambda e^{i 2 \pi r/\lambda}}{-i 2 \pi r}\left(\begin{array}{cc}S_2 & 0 \\ 0 & S_1\end{array}\right)\binom{E_{\| } ^{in}}{E_{\perp} ^{in}}, \quad 
    \label{Scatt_matrix}
\end{equation}

where $r$ is the radial direction and the subscripts $\|$ and $\perp$ indicate the parallel and perpendicular components of the field with respect to the scattering plane. Eq.~\ref{Scatt_matrix} dictates that $\left|S_2\right|^2$ and $\left|S_1\right|^2$ are proportional to the intensity scattered when the laser beam is fully polarized along the parallel or perpendicular direction, respectively (i.e. when $I_{\| } ^{in}$ =$I_{0}$ or $I_{\perp } ^{in}$=$I_{0}$). These matrix elements describe the visibility as $\mathcal{V}=(\left|S_2\right|^2-\left|S_1\right|^2)/(\left|S_1\right|^2+\left|S_2\right|^2)$. The sign of visibility indicates which orthonormal component of the electric field leads to a maximum or a minimum of the scattered intensity. 

We computed the elements of the scattering matrix and the visibility using the Mie formalism and the complex refractive index of the SCO NPs (see ellipsometry characterization in the Supporting Information). Fig.~\ref{fig3}(e) illustrates the calculated visibility as a function of the SCO NP size for three laser wavelengths, with a period of $\lambda$/2 and an amplitude that strongly depends on the refractive index. The visibilities obtained from the experiments correspond to the horizontal lines in Fig.~\ref{fig3}(e), where the intersections with numerical calculations provide nine possible radii smaller than 1\,$\mu$m. Thus, we employ three laser wavelengths to uniquely determine the size of the trapped particles. We allocate a Gaussian distribution centered at each intersection and calculate the convolution integral of the closest distributions at three wavelengths, which leads to a particle size of 315\,nm (see the vertical line in Fig.~\ref{fig3}(e)). This value sets the initial particle size at ambient pressure in the absence of laser excitation, which combined with new scattering experiments at 642\,nm is used to infer the subsequent particle's size change for increasing laser-excitation intensities. Fig.~\ref{fig3}(f) presents the visibility calculation at 642\,nm (dashed-red line) along with experimental values (markers) acquired upon laser excitation, alternating increasing and decreasing laser powers. Overall, an increase in laser power leads to a higher temperature for the particle, which in turn induces a spin transition and a decrease in the acquired scattering visibility. Herein, the laser-induced spin transition from the initial low- to the high-spin state of laser-heated particles leads to a size increase as expected.

Fig.~\ref{fig3}(g) provides the relative particle size change as a function of the laser excitation power, performed by cycling the laser power up and down. The analysis of these size–power curves reveals slightly asymmetric yet reproducible excitation-size trajectories in different NPs. The relative sizes are calculated by dividing the particle sizes at specific laser powers by the initial size in the absence of laser excitation. Our data reveal that increased excitation power induces a larger visibility shift (Fig.~\ref{fig3}(f)) and a larger particle size modification in Fig.~\ref{fig3}(g). The increase from a normalized relative size of 1.00 to the observed maximum of $\sim 1.06$ in Fig.~\ref{fig3}(g) corresponds to a laser-induced expansion of approximately $5 \pm 1\%$ in volume.

This abrupt light-induced size increase is consistent with typical values measured in SCO composite samples\cite{Molnr2017,dirtu2010insights,grosjean20111,yang2024room}. Ellipsometry measurements (Fig.~S5c) further support this interpretation, showing negligible thermal-induced thickness variations below and above the transition temperature range. This observation is supported by independent Raman spectroscopy measurements on non-trapped SCO NPs, which identify the HS state at 6~mW (dashed vertical line in Fig.~\ref{fig3}(g); see Raman measurements in Fig.~S4). This power range also aligns well with the thermal bistability of the Fe(II)–triazole coordination polymer and confirms the robustness of the spin transition under contact-free optical excitation.

\subsection*{Pressure modulation experiments on trapped SCO NPs}

\begin{figure}[H]
\centering
\includegraphics[width=\linewidth]{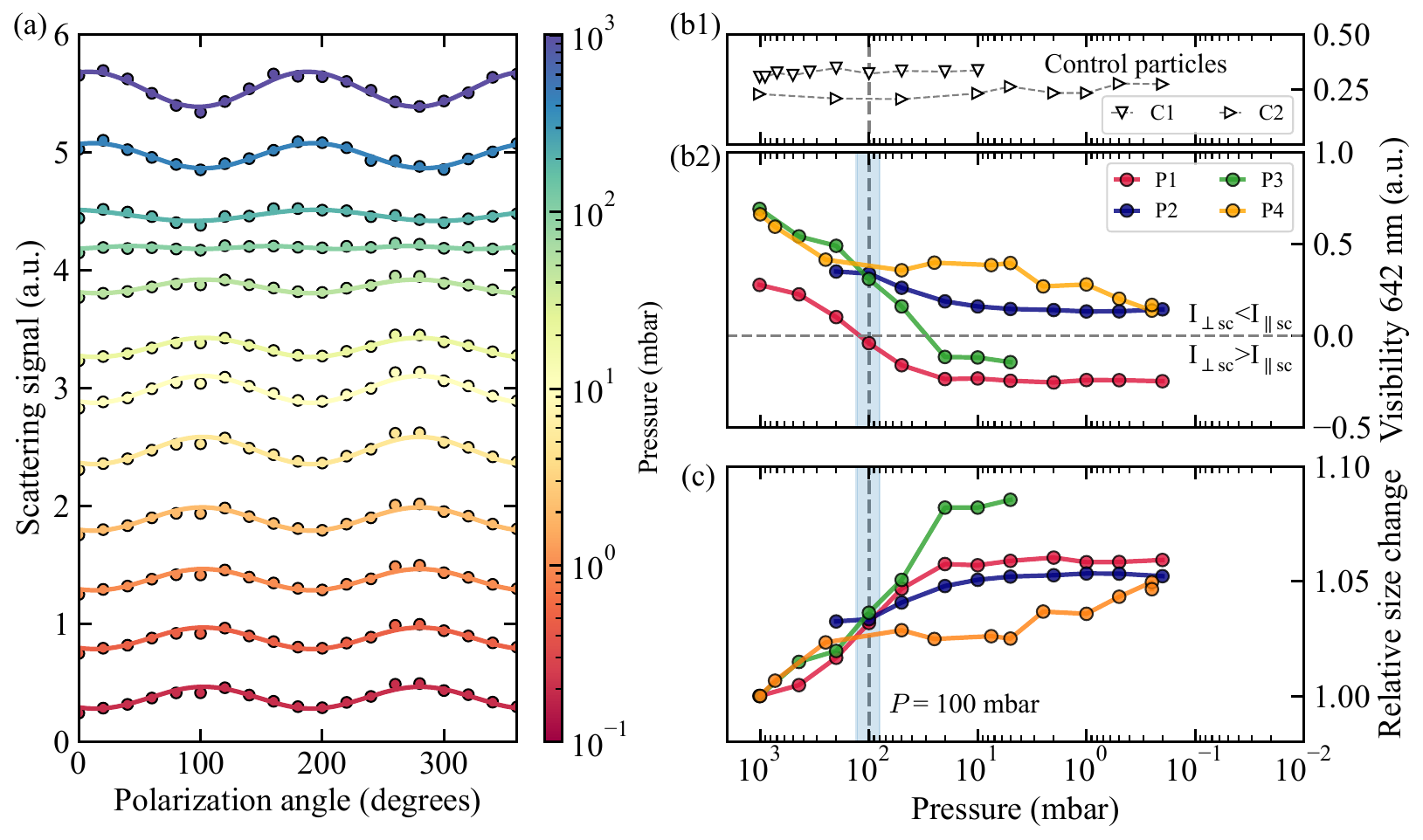}

\caption{\textbf{Pressure-dependent scattering of levitating SCO nanoparticles.} 
(a) Experimental polarization-resolved scattering at $\lambda_{\mathrm{probe}} = 642$~nm for a single trapped $\mathrm{Fe(NH_2trz)_3(NO_3)_2}$ nanoparticle while the ambient pressure is reduced from 1013 to 0.1~mbar. Curves are vertically offset for clarity and colored according to pressure (log scale). 
(b) Experimental visibility $V$ at 642~nm for six isolated particles as a function of ambient pressure. Their initial average NP size at ambient pressure is 280$\pm$4\,nm.
(b1) Control particles (C1, C2) introduced in the high-spin state exhibit pressure-independent $V$. 
(b2) Particles initially in the low-spin state show a monotonic decrease in $V$ that saturates below $\sim 10$~mbar. The vertical dashed line marks the approximate crossover pressure, and the horizontal dashed line separates $I_{\mathrm{Lsc}} < I_{\mathrm{\parallel sc}}$ ($V > 0$) from $I_{\mathrm{Lsc}} > I_{\mathrm{\parallel sc}}$ ($V < 0$). 
(c) Relative particle size change, extracted using the multi-wavelength visibility method, as a function of ambient pressure.}
\label{fig4}
\end{figure}

Next, we examine how reducing the ambient pressure affects levitating SCO particles. Figure~\ref{fig4}(a) shows representative 642~nm polarization scans as the pressure is lowered from 1013 to 0.1~mbar. The visibility $\mathcal{V}$ is positive at ambient pressure, flattens near $\sim 10^2$~mbar, and becomes negative at lower pressures, with sinusoidal amplitude recovering as pressure approaches a few mbar. The sign change and the evolution of the amplitude depend on the initial particle size and refractive index, consistent with the Mie calculations (cf. Figs.~\ref{fig3}(e)--(f)). We tracked $\mathcal{V}(P)$ for six particles (Fig.~\ref{fig4}(b)) with an initial average NP size at ambient pressure of 280±4 nm. The control particles introduced in the HS state (Fig.~\ref{fig4}(b1)) show pressure-insensitive visibilities, while the particles starting in the LS state exhibit a monotonic decrease of $\mathcal{V}$ that plateaus below $\sim 10$~mbar (Fig.~\ref{fig4}(b2)).

Using the multi-wavelength visibility method (Figs.~\ref{fig3}(e)--(f)), we converted these data to relative size changes (Fig.~\ref{fig4}(c)). Reducing the pressure from ambient pressure to $10^{-1}$~mbar produces a systematic decrease, and ultimately a sign change, in visibility, together with a 4--6\% increase in the inferred particle size. The most plausible origin of this ''vacuum-induced'' dilatation is dehydration of the Fe(II)--triazole coordination polymer: under vacuum the particles lose both weakly physisorbed water at the surface and a fraction of lattice (structural) water. Removal of these molecules (i) releases surface-induced compressive stress and capillary forces that clamp the lattice, and (ii) weakens the hydrogen-bond network that links $\mathrm{NH_2}$ groups and $\mathrm{NO_3^-}$ anions to the triazole chains. The latter reduces the effective ligand-field splitting at Fe(II), tipping the LS$\rightarrow$HS balance toward the expanded HS geometry and producing the observed volumetric increase through elastic cooperativity. Notably, because dehydration can modify the local crystal packing and hydrogen-bond topology, this mechanism may impart a partial or even non-reversible bias towards the HS state. The visibility ``turning point'' near $\sim10^2$~mbar is consistent with a threshold at which the desorption rate of water/solvent exceeds readsorption; it also aligns with observations on related Fe--triazole frameworks where depressurization into the $10^2$~mbar regime biases the equilibrium toward HS.\cite{SMIT20001697} Secondary contributions---reduced gas-phase cooling and damping at low pressure, which allow modest optical dissipation to raise the steady-state temperature---could potentially assist the shift but are insufficient to account for the magnitude on their own. The control particles prepared in the HS state show pressure-insensitive visibilities (Fig.~4(b1)), and the purely thermal expansion expected within a fixed spin state is far below the measured $5 \pm 1\%$, both observations reinforcing a dehydration-driven LS$\rightarrow$HS conversion. The particle-to-particle spread in the apparent crossover pressure (Fig.~4(b2)--(c)) is readily ascribed to variations in size, defect density/tilt boundaries, initial water loading, and net charge, which modulate both desorption kinetics and the local spin-state equilibrium.

\section*{CONCLUSIONS}

We have established a substrate-free platform to manipulate and interrogate spin crossover at the single-nanoparticle level. By coupling a quadrupole Paul trap with a multi-spectral, polarization-resolved scattering microscope, we isolate individual Fe(NH$_2$trz)$_3$(NO$_3$)$_2$ nanoparticles, excite them optically, and read out their state in real time while independently tuning the surrounding pressure. This configuration provides three key capabilities that have been elusive to date: (i) contact-free confinement that preserves the intrinsic response of the nanoparticle, (ii) quantitative, three-dimensional metrology of size and refractive-index--dependent scattering, and (iii) precise environmental control (laser intensity, pressure) at the level of a single particle.

Using this platform, we resolve light-driven switching between spin states with reversible opto-volumetric expansions up to $\sim$ 6\% and well-defined power thresholds. Independent pressure modulation yields a comparable expansion, and control measurements on particles prepared in the HS state remain insensitive to pressure. We attribute the observed ``vacuum-induced'' dilatation to dehydration of the Fe(II)–triazole coordination polymer, which can reduce the effective ligand-field splitting in Fe(II), shifting the LS$\leftrightarrow$HS equilibrium towards the expanded lattice that produces the observed volumetric increase.

Beyond providing a clear, mechanistic picture of single-particle switching, these results define operational windows such as intensity, pressure, and temperature, for integrating SCO NPs into nanophotonic circuitry. The ability to deterministically toggle and read out the spin state of an individual particle, free from substrate artifacts, points to energy-efficient optical switches, nonvolatile elements for dense data storage, and nanoscale sensors whose transduction rests on large, reversible volumetric and refractive-index changes.

We present a general methodology that can be extended to other SCO materials and molecular phase-change systems, to smaller particle sizes, and to architectures that couple levitated particles to optical cavities or integrated resonators for enhanced readout and feedback control. Combining the present approach with ultrafast excitation and on-chip trapping will enable time-resolved studies of cooperative dynamics, defect motion, and fatigue at the single-crystal level. Altogether, our work sets the stage for rational design and deployment of molecular phase-change building blocks in reconfigurable and ultralow-power nanophotonics.

\section*{EXPERIMENTAL METHODS}

\subsection*{SCO nanoparticle synthesis and characterization}

Synthetic protocols. All chemical reagents were purchased and used without further purification. Tetraethyl orthosilicate 98$\%$ (Sigma-Aldrich), Triton X-100 (Sigma-Aldrich), ascorbic acid (Sigma-Aldrich), 4-Amino-4H-1,2,4-triazole (Sigma-Aldrich), iron tetrafluoroborate hexahydrate (Sigma-Aldrich), barium nitrate (Sigma-Aldrich), Silicon Elastomer Sylgard 18 kit (Sigma-Aldrich), polymethyl metacrylate Mw 35000 (Sigma-Aldrich), n-hexanol (Sigma-Aldrich), cyclohexane (Sigma-Aldrich), ultra-pure water (18.2 M$\Omega$), ethanol absolute extra dry 99.5$\%$ (AcroSeal) and acetone HPLC grade (Scharlau).  

We synthesized Fe(NH$_2$trz)3(NO$_3$)$_2$ SCO NPs following the reverse micelle protocol, which consists of blending two separate microemulsions of the metal, Fe$^{2+}$ and the ligand 1,2,4-Amino Triazole. This protocol allowed us to control the size of the produced nanoobects by fine-tuning key parameters of the reaction.\cite{regueiro2024unlocking} In the IR spectra, we observed the characteristic stretching vibrations of the amine-triazole group around 1500 cm$^{-1}$ for C=C and N=N from the triazole, and 3100–3300 cm$^{-1}$ range corresponding to NH$_2$ and N-H groups of the amine (Fig. SX). XRD analysis showed that the material is highly crystalline and consists of a single phase, which matches well the theoretical pattern for the LS. NPs size distributions in solution were determined in ethanol (0.1\,mg/ml) suspensions by DLS using a Zetasizer ZS (Malvern Instrument, UK). Transmission Electron Microscopy studies were carried Technai G2 F20 microscope operating at 200 kV. Samples were prepared by dropping suspensions on lacey formvar/carbon copper grids (300 mesh). The real size distribution was determined by user-assisted counting of TEM images using ImageJ software. Attenuated total reflectance Fourier-transform infrared spectra were collected in an ALPHA II FTIR Spectrometer (Bruker) in the 4000-400 cm$^{-1}$ range without KBr pellets. Powder-X-ray Diffraction measurements were carried out in PANanlytical Empyrean diffractometer using Cu K$\alpha$ radiation (Cu K$\alpha$ = 1.5418 $\times10^{-10}$\,m) with a PIXcel detector, operating at 40 mA and 45 kV. Profiles were collected in 2$^\circ$ $<$ 2$\theta$ $<$ 45$^\circ$ range with a step size of 0.013$^\circ$. UV-vis absorption spectra were recorded on a Jasco V-670 spectrophotometer in baseline mode from 400 to 800\,nm range, using Thermo Scientific 96-well UV microplates containers. Magnetic data were collected with a Quantum Design MPMS XL-5 susceptometer equipped with a SQUID sensor. DC FC magnetization measurements were performed under an applied magnetic field of 100 Oe at 1Kmin$^{-1}$ scan rate in the temperature range between 100-400 K.

\subsection*{Trap and electrospray systems}
The trap setup combines an electrospray system, a vacuum chamber and a linear Paul trap. The Paul trap is built using four parallel gold-coated rods arranged to form a square of side L=8\,mm ($L_{rods}$ = 12\,cm and $r_{rods}$ = 3\,mm). The shape and size of the rods provides optical access to trapped particles while the so-produced electric field can be approximated to a field generated using hyperbolic rods.\cite{https://doi.org/10.1002/rcm.469,10.1116/1.1316304,Hernandez-Rueda:19} Rods placed at diagonally opposite vertices are attached to an amplifier that provides an AC voltage (U$_{pp}$=600\,V and $\Omega$=3.0\,kHz), while the other pair of rods is connected to a DC voltage (U=0-5\,V). In this configuration, our setup generates a time-varying electric quadrupole potential, which is able to trap single nano- to micron-sized particles following the so-called Mathieu equations.\cite{https://doi.org/10.1002/anie.199007391} The SCO particles are introduced in between the rods of the trap by means of an electro-spray system, which employs a syringe with a metallic needle that is set to 2900\,V. Such arrangement is placed close to the grounded trap´s entrance plate, thus originating a spray of charged particles with a conical shape (i.e. Taylor cone). The syringe contains a mix of approximately 20\,$\%$ powder of SCO particles and 80\,$\%$ of spectrally pure ethanol. Once the trapped particles are stable, we lock their position along the trap axis by activating the segmented rods. Subsequently, we activate the vacuum pump to gradually extract the air inside the transparent vacuum chamber.

\subsection{Polarization-resolved scattering microscopy}
The automated microscopy and polarimetry systems allow us to record snapshots of the scattered laser light intensity of levitated particles as a function of the laser polarization angle as shown in Fig.~\ref{fig3} (a). The polarimetry system uses a multiplexer to combine three linearly polarized pigtailed diode lasers with wavelengths 642\,nm, 785\,nm and 852\,nm. We employ these lasers to illuminate the trapped particles in order to acquire their scattering response and use it to characterize their size.\cite{Hernandez-Rueda:19} During the automated experiments, the laser polarization angle is set using an achromatic lambda-half wave plate (350-850\,nm, Thorlabs) mounted on a motorized rotation stage (PRM1/MZ8, Thorlabs). For each polarization angle, we collect 5-to-10 scattering micrographs using the in-situ microscope. The microscope employs a tube lens (f = 50\,mm) and a CCD camera (Zelux, Thorlabs) to record images of the scattering originated at the illuminated trapped particles at a 90 $^{\circ}$ angle with respect to the laser propagation direction. Each image is integrated along the $y$-axis and fitted using a Lorenztian function to extract a background-free scattering signal as shown in Fig.~\ref{fig3}(b). For every polarization angle, we average the scattering signal provided by the set of snapshots, yielding the scattering as a function of the laser probe (642\,nm) polarization angle upon different laser excitation conditions (488\,nm). Fig.~\ref{fig2}(c) presents an example of such curves for a trapped particle at ambient pressure (1013 mbar). In this case, the probe laser emits at $\lambda_p=642$\,m. However, we measure the scattering signal at three wavelengths at 642\,nm, 785\,nm and 852\,nm in order to unambiguously extract the particle size. To this end sinusoidal fits to the polarization-resolved scattering signal aid to extract the visibility at three wavelength, which is then used to infer the particle size as we explain in the following. 

We run two types of experiments where we inspect the scattering response of trapped particles as a function of pressure and moderate laser pump intensities. For the first experiments, trapped particles are investigated at specific pressure values inside the vacuum chamber in the 0.01-1013 mbar range. To this end, we use a coarse-and-fine valve system (Edwards) that connects the vacuum chamber of the trap to a scroll pump (Edwards). The pressure inside the chamber is monitored using a gauge (TPG50 and TPG51, Balzers). In order to prevent the particles from exiting the trap during subsequent air exhaust procedures, we critically employ the segments of the trap to generate an additional electric potential that retains the particles along the axis of the trap. For the laser excitation experiments, we employ a  CW blue pump laser at 488\,nm (LP400-SF20G), which provides powers in the range of 30\,\textnormal{$\mu$}W to 100\,mW. The laser diode is embedded in a temperature and impedance control mount (LDM9LP, Thorlabs) connected to a power supply (ITC4000, Thorlabs). The laser is guided via fiber optics to a port that uses a collimator to direct the beam to a microscope objective (Mitutoyo, Plan-Apochromat objective, 10$\times$, NA=0.28, working distance 34\,mm). The laser fiber and objective lens are mounted on a three-dimensional translation stage that allows us to optimize the beam-particle alignment in real time by maximizing the scattered light collected by the microscope. Details of the laser beam characterization procedure can be found in the Supporting Information. To estimate the precise laser fluence that illuminates the SCO nanoparticle surface, we consider the formula that accounts for the energy-on-particle that is detailed elsewehere.\cite{hernandez2022early,meijer2022nanosecond} We filter out the laser excitation wavelength employing a long-pass filter (FGL530M, Thorlabs) in order to acquire 405\,nm-free snapshots during the polarimetry experiments.

\section*{Supporting information}

The supporting information contains extended details of the experimental methods. We provide IR, PXRD and DLS measurements, the characterization of the complex refractive index of SCO particles using ellipsometry, and laser beam procedure details.

\section*{Funding}

E.P.-C acknowledges funding from Generalitat Valenciana (Grant No. SEJIGENT/2021/039) and Agencia Estatal de Investigación of Ministerio de Ciencia e Innovacion (PID2021-128442NA-I00 and CNS2024-154922 funded by MICIU/AEI /10.13039/501100011033). J. H-R. acknowledges funding from the Programa de Atracción de Talento de la Comunidad Autónoma de Madrid, Modalidad 1, Project 2020-T1/IND-19951. The authors acknowledge funding from the European Commission (Pathfinder-4D-NMR 101099676), the Spanish Ministerio de Ciencia e Innovación (Unit of Excellence “Maria de Maeztu” grant CEX2024-001467-M, PID2023-149309OB-I00 and PID2020–117264GB-I00 funded by MICIU/AEI/10. 13039/501100011033). This study forms part of the Advanced Materials program and was supported by MCIU with funding from the European Union NextGenerationEU (PRTR-C17.I1) and by Generalitat Valenciana (MFA/2022/050 and MFA/2022/025). The authors also thank the Generalitat Valenciana PROMETEO Program (CIPROM/2024/51).

\begin{acknowledgement}

The authors thank Prof. Dries van Oosten for his contributions to the trap design and polarimetry conceptualization and for lending the trap equipment to J. H-R, Dante Killian and Paul Jurrius for their contributions to the fabrication of the trap system. The authors thank Dr. Teodora Ivanova Angelova for her contributions to ellipsometry characterization. The authors thank Prof. Rosa Weigand for lending the vacuum pump used during the experiments.

\end{acknowledgement}


\end{document}


\newpage

\section{S1. Zeta-potential measurements }

\section*{Zeta-potential characterization of [Fe(NH$_2$trz)$_3$](NO$_3$)$_2$ NPs}

\begin{figure}
    \centering
    \includegraphics[width=0.6\linewidth]{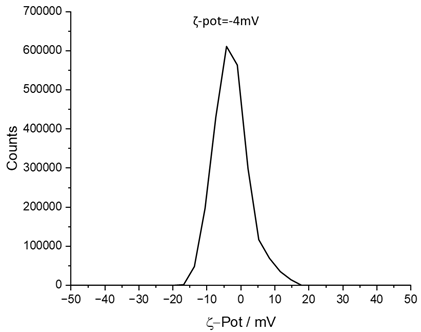}
    \caption{$\zeta$-potential measurements.}
    \label{S1}
\end{figure}

To confirm the presence of surface charges in isolated SCO nanoparticles, we performed $\zeta$-potential measurements (Fig.~\ref{S1}). The nanoparticles exhibit a $\zeta$-potential of approximately $-4$~mV, consistent with the partial surface charging expected from the presence of multiple NO$_3^-$ counterions in the structure. This observation supports the suitability of [Fe(NH$_2$trz)$_3$](NO$_3$)$_2$ as a platform for the preparation of voltage-responsive SCO nanomaterials.

\newpage

\section{S2. Dynamic Light Scattering measurements}

\begin{figure}[H]
    \centering
    \includegraphics[width=0.7\linewidth]{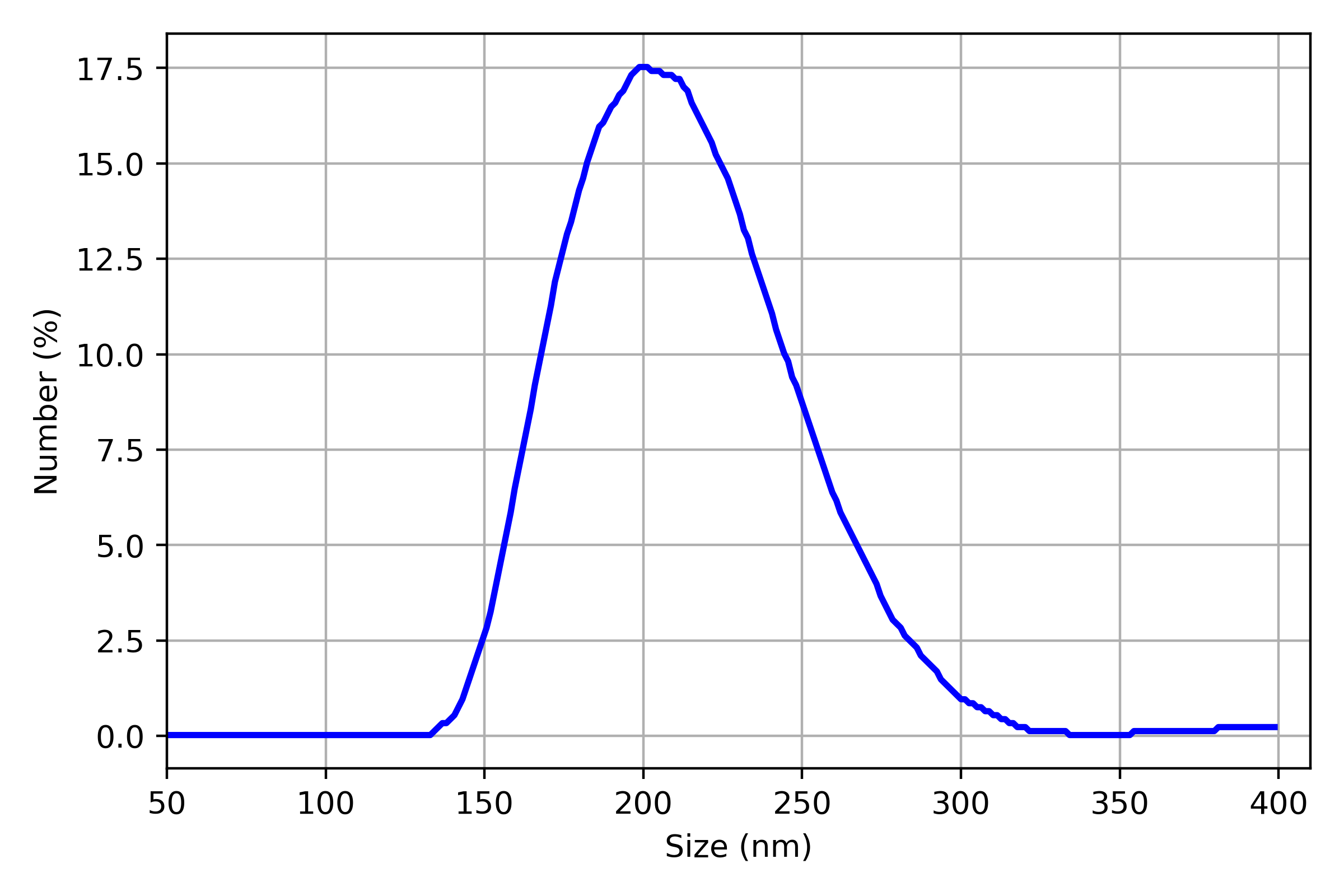}
    \caption{Dynamic Light Scattering measurements}
    \label{S2}
\end{figure}

In Fig.~\ref{S2}, we plot a dynamic light scattering (DLS) measurement of our [Fe(NH$_2$trz)3(NO$_3$)$_2$] nanoparticles. The measurements were performed at room temperature using a Zetasizer ZS (Malvern Instrument, UK). The size distribution curve reveals a well-defined peak centered around 200\,nm, indicating distinct populations of nanoparticle as small as 130\,nm and as large as 400\,nm.

\newpage
\section{S3. Infrared and X-Ray Diffraction measurements}

\begin{figure}[H]
    \centering
    \includegraphics[width=\linewidth]{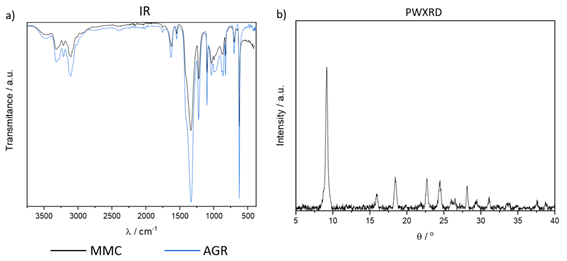}
    \caption{a) Infrared and b) Powder X-Ray Diffraction measurements}
    \label{S3}
\end{figure}

In Fig.~\ref{S3} we illustrate (a) infrared (IR) spectra and (b) powder X-ray diffraction (PXRD) pattern of the Fe(II)–triazole coordination polymer before (MMC, black) and after (AGR, blue) the nanoparticle fabrication process. The IR spectra confirm the preservation of the characteristic vibrational modes of the triazole-based framework, with only minor shifts observed, indicating retention of the molecular structure. The PXRD pattern exhibits sharp and well-defined diffraction peaks, consistent with a crystalline structure, and in agreement with previously reported patterns for this compound class. These results validate the chemical integrity and crystallinity of the SCO material throughout the fabrication and processing steps.

\newpage

\section{S4. Measured Raman spectra}

Raman spectroscopic characterization was carried out using an HR Evolution confocal Raman microscope (Horiba) with a laser spot size of approximately 1 $\mu$m (Olympus LMPlanFl 50x LWD, NA 0.50) in backscattering geometry. A laser with an excitation wavelength of $\lambda_{exc}$= 473 nm was used to induce the spin state transition, while $\lambda_{exc}$= 633 nm served as the probe for spin state detection. The incident laser power for both lasers was maintained at 25\% of the maximum  which corresponds to 6 mW for the blue laser (473 nm). A diffraction grating with 600 grooves/mm was used.

\begin{figure}
    \centering
    \includegraphics[width=0.7\linewidth]{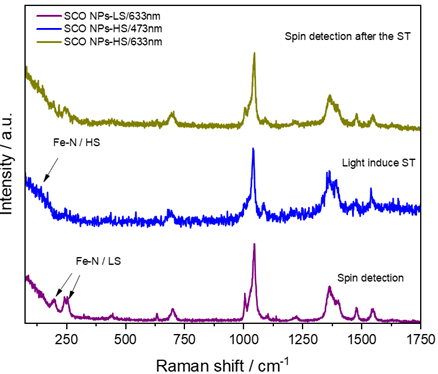}
    \caption{Raman spectra of the SCO NPs measured with a 633 nm excitation wavelength (bottom). Spectra recorded with a 473 nm excitation wavelength for the light-induced spin transition (middle). Finally, the vibrational signature of the NPs after the spin transition, measured again with a 633 nm excitation wavelength (top).}
    \label{S4}
\end{figure}

\newpage
\section{S5. Elipsometry measurements}

Ellipsometric analysis of the nanocomposite thin film. The sample, consisting of SCO/PMMA, was spin-coated onto a silicon substrate, yielding a uniform and optically smooth film. The film was characterized with a UV-visible-NIR SENTEC spectroscopic ellipsometer equipped with a programmable heating stage (from 300~K to 350~K). Spectroscopic ellipsometry enabled us to determine the complex refractive index $(n + i k)$ of the Fe(NH$_2$trz)$_3$(NO$_3$)$_2$ compound embedded in the polymer matrix.

\begin{figure}[H]
    \centering
    \includegraphics[width=.75\linewidth]{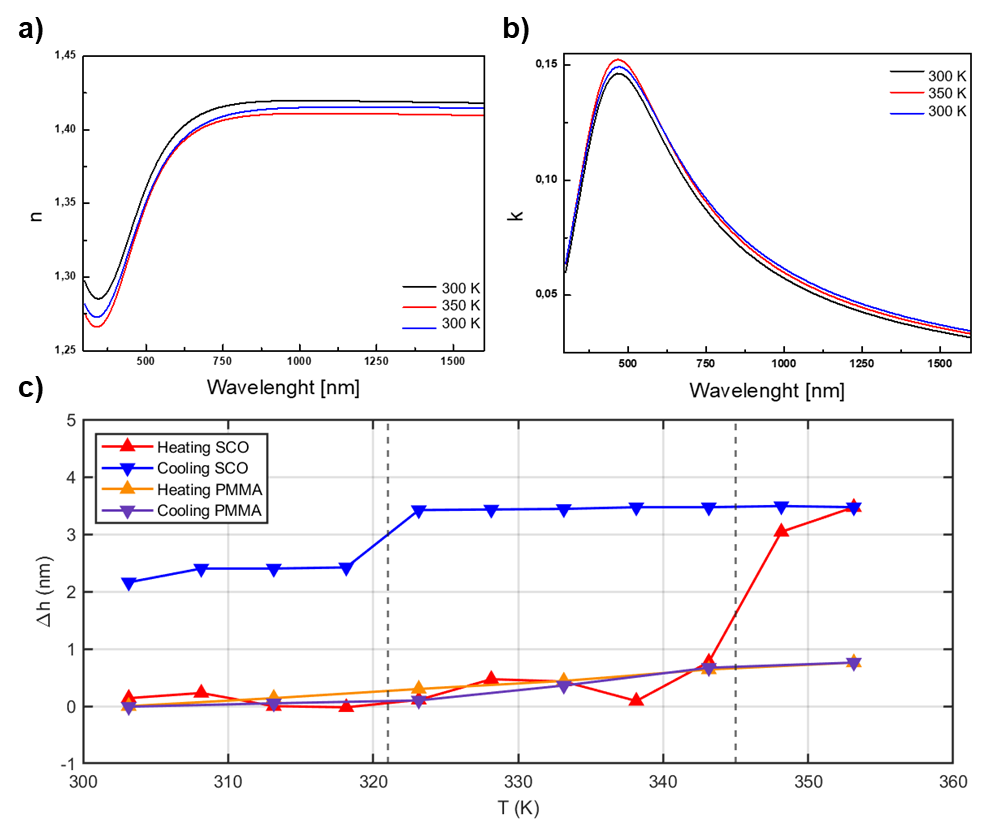}
    \caption{Ellipsometry data. (a) Refractive index ($n$) and (b) extinction coefficient ($k$) as a function of wavelength at different temperatures (300~K, 350~K, and 300~K after cooling). (c) Temperature-dependent thickness variation ($\Delta h$) during heating and cooling cycles for SCO/PMMA and PMMA films.
}
    \label{S5}
\end{figure}

The results from the SCO plus PMMA resist layer were fitted from ellipsometry data using Tauc--Lorentz oscillators\cite{Jellison1996ParameterizationOT}, allowing an accurate description of both the dispersion of the refractive index ($n$, Figure 5a) and the extinction coefficient ($k$, Figure 5b), as a function of wavelength at different temperatures (300~K, 350~K, 300~K cooling). 
Figure 5c shows the temperature-dependent thickness variation ($\Delta h$) of the SCO/PMMA and PMMA films. 
For the SCO/PMMA layer, a  thermal hysteresis is observed for the SCO/PMMA were the PMMA reference displays a linear 
increase in thickness (below 1~nm) across the entire temperature range.

\newpage
\section{S6. Pump laser beam size characterization inside the trap}

The calibration of the size of the pump laser beam that was focused on the trapped particles in the trap was carried out at a wavelength of 488\,nm. The focused laser beam was scanned laterally across a static trapped particle, and the corresponding scattering signal was recorded as a function of the beam position along the x-axis parallel to the trap axis. The experimental profiles were analyzed by fitting a Gaussian function to reliably determine the spatial beam distribution. This procedure was repeated multiple times to ensure reproducibility and reduce statistical uncertainty.\cite{maestrohernandez2025}
